\documentclass[conference]{IEEEtran}
\IEEEoverridecommandlockouts
\usepackage{cite}
\usepackage{booktabs}
\usepackage{amsmath,amssymb,amsfonts}
\usepackage{algorithmic}
\usepackage{graphicx}
\usepackage{textcomp}
\usepackage{xcolor}
\def\BibTeX{{\rm B\kern-.05em{\sc i\kern-.025em b}\kern-.08em
    T\kern-.1667em\lower.7ex\hbox{E}\kern-.125emX}}
\begin{document}

\title{A Scalable AI-Powered System for Explainable Machine Learning Pipelines in Brain Tumor\\
\thanks{\\$^{*}$These authors contributed equally as co-last authors.\\
$^{1}$This author is also affiliated with Fondazione IRCCS istituto Neurologico Carlo Besta}
}

\author{\IEEEauthorblockN{1\textsuperscript{st} Yin Lin}
\IEEEauthorblockA{\textit{DEIB} \\
\textit{Polytechnic University of Milan}\\
Milan, Italy \\
yin.lin@polimi.it}
~\\
\and
\IEEEauthorblockN{2\textsuperscript{nd} Elena De Martin}
\IEEEauthorblockA{\textit{Health Department} \\
\textit{Fondazione IRCCS Istituto Neurologico}\\
\textit{Carlo Besta}\\
Milan, Italy \\
elena.demartin@istituto-besta.it}
~\\
\and
\IEEEauthorblockN{3\textsuperscript{rd} Giacomo Conte}
\IEEEauthorblockA{\textit{DEIB} \\
\textit{Polytechnic University of Milan}\\
Milan, Italy \\
giacomo.conte@mail.polimi.it}
~\\
\and
\IEEEauthorblockN{4\textsuperscript{th} Domenico Aquino}
\IEEEauthorblockA{\textit{Neuroradiology Unit} \\
\textit{Fondazione IRCCS Istituto Neurologico}\\
\textit{Carlo Besta}\\
Milan, Italy \\
domenico.aquino@istituto-besta.it}
~\\
\and
\IEEEauthorblockN{5\textsuperscript{th} Cristiana Pedone}
\IEEEauthorblockA{\textit{Radiotherapy Unit} \\
\textit{Fondazione IRCCS Istituto Neurologico}\\
\textit{Carlo Besta}\\
Milan, Italy \\
cristiana.pedone@istituto-besta.it}
~\\
\and
\IEEEauthorblockN{6\textsuperscript{th} Alberto Redaelli}
\IEEEauthorblockA{\textit{DEIB} \\
\textit{Polytechnic University of Milan}\\
Milan, Italy \\
alberto.redaelli@polimi.it}
\and
\IEEEauthorblockN{7\textsuperscript{th} Riccardo Barbieri}
\IEEEauthorblockA{\textit{DEIB} \\
\textit{Polytechnic University of Milan}\\
Milan, Italy \\
riccardo.barbieri@polimi.it}
~\\
\and
\IEEEauthorblockN{8\textsuperscript{th} Laura Fariselli$^{*}$}
\IEEEauthorblockA{\textit{Radiotherapy Unit} \\
\textit{Fondazione IRCCS Istituto Neurologico}\\
\textit{Carlo Besta}\\
Milan, Italy \\
laura.fariselli@istituto-besta.it}
~\\
\and
\IEEEauthorblockN{9\textsuperscript{th} Simona Ferrante$^{*}$$^{1}$}
\IEEEauthorblockA{\textit{DEIB} \\
\textit{Polytechnic University of Milan}\\
Milan, Italy \\
simona.ferrante@polimi.it}
}

\maketitle

\begin{abstract}
Artificial intelligence and radiomics are increasingly used in brain tumor research, yet their translation into clinical practice remains limited by fragmented workflows, poor transparency, and weak integration with end users’ needs. We present the first version of a scalable web-based visual analytics system designed to support radiomics-driven machine learning inference in neuro-oncology. The platform integrates three core functions within a single interface: cohort management from structured clinical tables, radiomic feature extraction from medical images and segmentation masks, and guarded inference with pre-trained machine learning models. The system was developed through an iterative user-centred design process and evaluated on both a public glioblastoma dataset and a proprietary clinical cohort. A key contribution is the explicit exposure of intermediate workflow artifacts, which improves traceability, interpretability, and responsible use of AI. By combining portability, inspectability, and deployment simplicity, the proposed framework offers a practical foundation for clinically oriented AI applications in brain tumor analysis.
\end{abstract}


\section{Introduction}
Brain tumors represent one of the oncological fields with the highest biological and clinical heterogeneity \cite{neftel2019integrative}. The WHO classification is associated with substantial prognostic differences: patients with WHO grade I meningioma generally show a 10-year overall survival of about 80–90\% \cite{champeaux2021nationwide}, whereas in glioblastoma the median overall survival is around 12 months \cite{stupp2005radiotherapy}. This marked variability, together with the increasing availability of multi-modal data (imaging, clinical, molecular, and laboratory variables), has fostered in recent years an intense development of artificial intelligence–based approaches with the aim of supporting therapy personalization \cite{ligero2025artificial} and risk stratification \cite{van2020radiomics,lin2025glioblastoma,lin2026lightweight}.

In the literature, the most widespread paradigm for brain tumors is the so-called “two-step” pipeline \cite{van2020radiomics,yuhandri2023improving}. In a first phase, biomedical images (e.g., MRI or CT) are segmented, manually or through automatic algorithms, to delineate tumor regions (ROI) on which to perform the extraction of quantitative characteristics; radiomics represents the most consolidated approach in this context \cite{lambin2012radiomics}. Subsequently, radiomic features are integrated with other information (clinical and, when available, biological) through data fusion strategies and used as input for machine learning or deep learning models, often targeting tasks such as outcome prediction (e.g., survival) \cite{kickingereder2016radiomic}, subtype classification \cite{abidin2019investigating}, or biomarker estimation \cite{arita2018lesion}.

Despite methodological progress, research attention on brain tumors has focused mainly on technical aspects such as pre-processing, model architecture design, training strategies, and explainability methods \cite{samala2024ai}. Conversely, less emphasis is devoted to operational transferability and integration into clinical workflows. This leads to a recurring critical issue: the lack of an effective “bridge” between those who develop models and those who must interpret and use them in the context of brain tumors. In this sense, the risk is that promising experimental solutions remain confined to laboratory prototypes, with limited impact on real-world clinical practice \cite{kocak2025widening}.

In light of these considerations, this work addresses the need to make the entire radiomics modeling process more accessible and usable in an applied setting. The main contributions are as follows:

\begin{itemize}
    \item Design and implementation of a scalable AI-enabled visual analytics system, intended to facilitate interaction between clinical stakeholders and engineering expertise, reducing the barrier to access analytical tools.
    \item Native integration of pre-trained machine learning models within the platform, enabling inference from user-provided data and timely delivery of results.
    \item End-to-end visualization of the modeling pipeline in an interactive, web-based manner, including feature extraction and interpretation of results (e.g., via SHAP), to increase transparency, traceability, and usability.
\end{itemize}

\section{Related Works}
In oncology, interest in visualization systems that support multidisciplinary clinical practice is steadily increasing \cite{boehm2025data}. The literature remains largely monitored: most dashboards are designed to track indicators, temporal trends, and outcomes, with limited emphasis on structured interaction between clinicians and the underlying analytical pipelines \cite{bucalon2022state,helminski2024development}. Within this stream, Mohindra et al. \cite{mohindra2024development} developed an EHR-integrated dashboard to support shared decision-making by organizing patient-reported outcomes (PROs) and longitudinal data into clinic-facing views. Perry et al. \cite{perry2022patient} highlighted real-world adoption constraints during implementation, indicating that dashboard availability alone does not ensure sustained use or stable data quality over time. In the molecular tumor board setting, Strantz et al. \cite{strantz2025empowering} introduced and discussed digital tools that aggregate clinical–molecular evidence to support collegial decision-making.

When imaging enters the workflow, systems tend to focus on two main tasks: labelling and segmentation. For labelling, Diaz-Pinto et al. \cite{diaz2024monai} proposed an AI-assisted interactive annotation framework that reduces manual burden and standardizes user–model iteration. For segmentation, established image computing platforms such as 3D Slicer \cite{fedorov20123d} provide an extensible ecosystem for visualization, segmentation, and quantitative analysis.

In brain tumors, the literature is even more limited. Sailunaz et al. \cite{sailunaz2023brain} proposed an interactive framework for detection and segmentation with a user interface and feedback mechanism to improve accuracy and trust. Severn et al. \cite{severn2022pipeline} presented a radiomics pipeline with explainability modules and result visualization, showing the feasibility of interpretable clinical reports. However the focus remains on viewing/segmentation rather than integrating the full learning chain into a single interactive system \cite{kikinis20113d,lee2025ai}.

A coherent contribution that implements and makes inspectable the entire radiomics ML pipeline for brain tumors such as preprocessing, feature extraction/selection, training, validation, explainability, reporting, and experiment comparison, within a single interactive system therefore remains unaddressed.

\begin{figure*}%% placement specifier
%% Use \includegraphics command to insert graphic files. Place graphics files in 
%% working directory.
\centering%% For centre alignment of image.
\includegraphics[width=15.5cm]{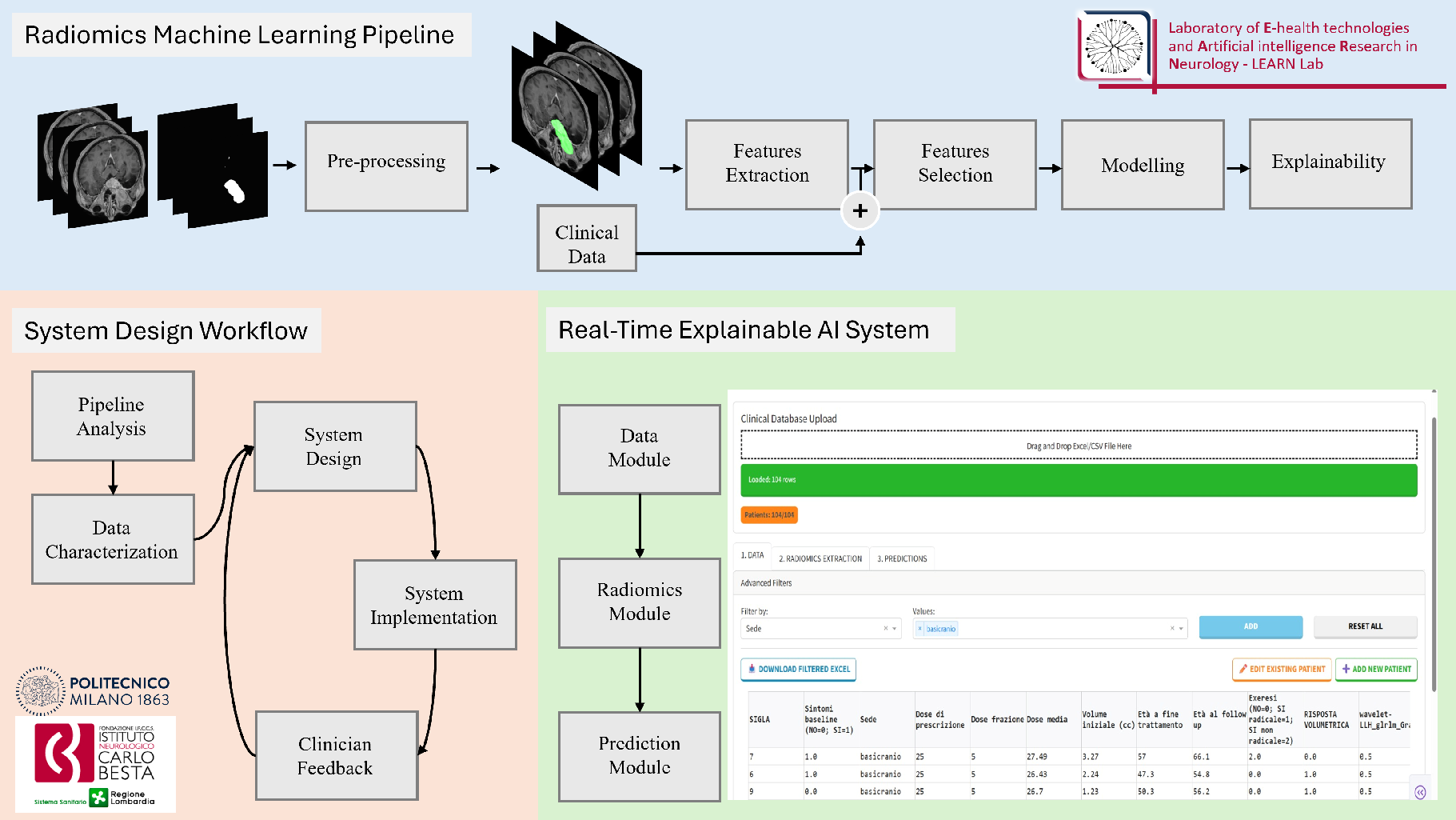}
%% Use \caption command for figure caption and label.
\caption{Overview of the proposed system. Upper panel: the complete radiomics machine learning pipeline for brain tumor analysis, including image pre-processing, radiomic feature extraction, integration with clinical data, feature selection, modeling, and explainability. Bottom-left panel: the system design workflow, structured as an iterative and user-centered process including pipeline analysis, data characterization, system design, implementation, and clinician feedback. Bottom-right panel: screenshot of the implemented system, showing the Data module with cohort filtering functionalities, and the three main integrated modules: Data, Radiomics, and Prediction.}\label{pipelines}
%% https://en.wikibooks.org/wiki/LaTeX/Importing_Graphics#Importing_external_graphics
\end{figure*}

\section{Methods}

\subsection{Data}
The system was implemented and tested on two data sources. The first dataset is BraTS 2020 \cite{menze2014multimodal,bakas2017advancing,bakas2018identifying}, widely used as a benchmark for segmentation and prediction tasks in neuro-oncology in glioblastoma. The second dataset is a proprietary clinical cohort collected at the Fondazione IRCCS Istituto Neurologico Carlo Besta (INCB), approved by the Institutional Ethics Committee (August 11, 2011; protocol 415/2011, Comitato Etico Regione Lombardia, section Fondazione IRCCS Istituto Neurologico).

The BraTS 2020 dataset includes 236 subjects with glioblastoma. For each subject, age is available as a clinical variable and four MRI sequences (T1, T1ce, T2, FLAIR), associated with the corresponding segmentation mask. The main cohort characteristics are reported in Table \ref{tab:brats2020}.

The INCB dataset includes 104 patients with skull base meningioma treated with CyberKnife. For each patient, essential clinical variables are available (e.g., sex, lesion location, baseline tumor volume, age at treatment, and prescription dose) and pre-treatment imaging data, consisting of T1-weighted MRI co-registered with CT images. Tumor segmentation is available for all volumes and were produced by INCB radiologists using semi-automatic software routinely employed for radiotherapy treatment planning. The main cohort characteristics are reported in Table \ref{tab:incb_besta}.

\begin{table}[t]
\caption{Clinical characteristics of the BraTS2020 cohort}
\label{tab:brats2020}
\small
\centering
\begin{tabular}{l c}
\toprule\toprule
\textbf{Characteristic} & \textbf{Median (Min-Max) or Number} \\
\midrule
Number of subjects & 236 \\
Age at diagnosis (years) & 61.5 (19.0--86.7) \\
Overall survival (days) & 370 (5--1767) \\
\bottomrule
\end{tabular}
\end{table}

\begin{table}[t]
\caption{Clinical and treatment characteristics of the INCB (Besta) cohort}
\label{tab:incb_besta}
\small
\centering
\resizebox{\columnwidth}{!}{%
\begin{tabular}{l c}
\toprule\toprule
\textbf{Characteristic} & \textbf{Median (Min-Max/\%) or Number} \\
\midrule
Number of subjects & 104 \\
Age at treatment (years) & 57.1 (19.8--86.8) \\
Sex (female) & 84 (80.8\%) \\
Tumour volume before treatment (cc) & 11.6 (0.8--57.4) \\
Mean dose (Gy) & 27.5 (21.9--30.8) \\
\midrule
Volumetric response: & \\
1) Progressive Response (PR) & 67 (64.4\%) \\
2) Stable Disease (SD) & 32 (30.8\%) \\
3) Progressive Disease (PD) & 5 (4.8\%) \\
\bottomrule
\end{tabular}%
}
\end{table}

\subsection{Design Process}
The system design process adopted an iterative, user-centered workflow, where pipeline analysis informed an initial prototype that was subsequently refined through clinician feedback loops \cite{shuldiner2023developing}. The overall pipeline is illustrated in the bottom-left of Figure \ref{pipelines}. The process is divided into five main steps.

\begin{enumerate}
    \item Analysis of the ML pipeline for brain tumor: We decomposed the standard pipeline, shown in the upper panel of Figure \ref{pipelines}, to identify the most relevant steps, both from an informational and an operational perspective, for the clinical user. The considered phases include: (i) extraction of radiomic features, (ii) concatenation of features (radiomic and clinical), (iii) a validation scheme with nested cross-validation, including feature selection, training, and testing, and (iv) explainability of results at the model level and at the single-patient level. This analysis allowed us to define which steps of the workflow should be made inspectable in the dashboard.
    \item Analysis of the source data formats: We characterized the input formats to ensure compatibility with existing workflows. MRI images and segmentation masks are represented in NIfTI format, while clinical variables are structured in tabular form and stored in CSV or XLSX files. This heterogeneity motivated an explicit separation between imaging management components and tabular data management components, while keeping an integration point at the level of the feature set and patient metadata.
    \item Logical design of the system components: Based on the workflow analysis, we defined the functional modules of the system along the chain “data, features, model, explanation”. From experimental practice, preprocessing emerged as time-consuming, strongly dependent on implementation choices, and not the main contribution of this work; for this reason, it was excluded from the system scope. Similarly, for the modeling part we assume that models are already trained and available as reusable artifacts. The dashboard is therefore designed to: (i) load and organize inputs and features, (ii) make validation and test results inspectable, and (iii) provide explanations and comparisons across experiments and subgroups.
    \item Implementation and deployment constraints: Deployment was constrained by the typical hardware resources observed in the clinical environment (limited CPU and RAM, outdated workstations). In this context, a standard architecture with a dedicated backend and persistent services was not well suited in terms of maintenance and portability. We therefore implemented the dashboard as a web app, prioritizing lightness and reproducibility. For the implementation we adopted the Dash library, which enables the delivery of an interactive web interface while keeping a simple and easily replicable execution model.
    \item Iteration with clinical feedback: Once a functional prototype was developed, the system was shown to clinicians and discussed in dedicated sessions to collect feedback on functionalities and visual design choices, with the goal of improving usability and alignment with real workflows. The process follows a closed loop “prototype, qualitative evaluation, revision”, oriented to applicability in the clinical setting.
\end{enumerate}
In this study we present the first version of the system (V0) resulting from this design process.

\section{Results: System implementation (V0)}
The first version of the system, illustrated in the bottom-right panel of Figure \ref{pipelines}, consists of three integrated modules: Data, Radiomics, and Prediction.

The Data module supports cohort ingestion from Excel or CSV files. Upon upload, the system automatically populates an interactive full-screen table, enabling immediate inspection of the cohort and available clinical variables. Cohort exploration is enabled through an advanced filtering component that supports multiple, simultaneous inclusion/exclusion criteria. Active criteria are shown as removable tags, while a real-time counter in the top-left corner reports the size of the current subset. The resulting filtered sub-cohort can be exported directly. The module also supports cohort extension: users can register a new subject via a structured form, which appends the record to the active database.

The Radiomics module provides dedicated upload zones for the MRI volume, the corresponding segmentation mask, and a PyRadiomics configuration file (\texttt{.yaml}). After feature extraction, the generated radiomic descriptors are displayed in a preview table to support rapid verification. A key functionality is the “Append to Database” operation: once selected, the system merges the newly extracted high-dimensional radiomic features into the active clinical database by matching the unique patient identifier. This step ensures that newly added subjects are immediately available for downstream AI-based inference without manual reconciliation across data sources.

The Prediction module enables on-demand inference using the embedded models developed in this project. Users select a patient identifier from a drop-down menu and request a real-time prediction of the target label. To reduce the risk of inappropriate use, the system implements explicit guardrails at inference time. If the user attempts to apply a model that is incompatible with the selected dataset, inference is blocked and no prediction is returned. The interface displays an orange warning banner that reports the mismatch, preventing potentially misleading outputs caused by an improperly applied model.

Overall, V0 provides an end-to-end operational layer that links cohort handling, radiomics extraction, and safeguarded prediction in a single dashboard, with each module designed to keep intermediate artifacts inspectable and actions auditable at the point of use.

\section{Discussion}
This study presents the first version of a visual analytics system designed to support radiomic machine learning pipelines in brain tumors. The contribution is situated in an area where the literature is still fragmented: many works stop at visualizing results or providing tools for segmentation, while systems that operationalise, within a single interface, the key steps of the radiomics–ML chain remain rare \cite{kikinis20113d,lee2025ai}. In this sense, the proposed tool helps fill a concrete gap by offering an environment that does not limit itself to the presentation of the final output, but instead organizes and makes the essential workflow artifacts inspectable.
A distinctive element is the transparency of the process. The system explicitly exposes the transition between cohort management (clinical variables), generation and integration of radiomic features, and inference through pre-trained models. This approach reduces the opacity that typically characterizes “black-box” adoption and facilitates the responsible use of AI in neuro-oncology, where application errors may generate unreliable outputs. The guardrails introduced in the prediction module respond to this critical issue by blocking inference under inappropriate conditions and making the problem evident to the user.
From a translational perspective, the implementation choice of a lightweight web app is coherent with realistic constraints of the clinical environment. The absence of specific hardware requirements facilitates integration into daily work, especially in contexts where clinical workstations have limited resources. From this perspective, the system prioritizes portability and reproducibility, while maintaining a deployment model compatible with existing infrastructures.
Relevant limitations are nevertheless present. First, despite the effort towards generalization with respect to heterogeneous sources, the current implementation is constrained to specific formats: tabular data in CSV/XLSX and imaging/masks in NIfTI. For broader adoption, compatibility must be extended to formats commonly used in clinical settings and hospital systems, in particular DICOM and related metadata management modalities. Second, the online extraction of radiomic features may be slow, especially on non-dedicated hardware. This requires optimisation of the process, for example through caching, controlled parallelisation, or asynchronous modes with state management and notifications. Third, the functional coverage is still incomplete. V0 represents a starting point and requires subsequent iterations guided by clinical feedback, both in terms of functions and in terms of interface.

\section{Conclusion}
We presented the first version of scalable AI-Powered System to support radiomics workflows and ML inference in brain tumors. The system combines cohorts management, radiomic feature extraction and fusion, and prediction with guardrails, promoting transparency and responsible use of AI. The next steps include extending support to the most common clinical formats, improving the performance of radiomic feature extraction, and continuing a refinement cycle with clinicians to increase applicability and functional completeness.

\vspace{12pt}

\end{document}